# Centralized Integrated Spectrum Sensing for Cognitive Radios

Dilip S. Aldar

¹K.B.P.College of Engineering & Polytechnic Satara
Email: dilip_aldar@rediffmail.com

**ABSTRACT**

Spectrum sensing is the challenge for cognitive radio design and implementation, which allows the secondary user to access the primary bands without interference with primary users. Cognitive radios should decide on the best spectrum band to meet the Quality of service requirements over all available spectrum bands. This paper investigates the integrated centralized spectrum sensing techniques in multipath fading environment and the performance was analyzed with energy detection and wavelet based sensing techniques for unknown signal.

*Keywords:* Cognitive Radio, Spectrum Sensing, Signal Detection, Primary User, Secondary User

## 1. INTRODUCTION

Observing that in some locations or at some times of day, 70 percent of the allocated spectrum may be sitting idle, the FCC [1] has recently recommended that significantly greater spectral efficiency could be realized by deploying wireless devices that can coexist with the primary users, generating minimal interference while taking advantage of the available resources. Thus, the discrepancy between spectrum allocation and spectrum use suggests that this spectrum shortage could be overcome by allowing more flexible usage of a spectrum. Flexibility would mean that radios could find and adapt to any immediate local spectrum availability.

Spectral efficiency plays an increasingly important role as future wireless communication systems will accommodate more and more users and high performance rich content services. Cognitive radio technologies have been proposed for lower priority secondary systems aiming at improving spectral efficiency by sensing the environment and then filling the discovered gaps of unused licensed spectrum with their own transmission [2]. In addition, cognitive radio can be used within licensed network to improve the spectral efficiency. The aim of cognitive radio is efficient use of natural resources, which include frequency, time, and transmitted energy. Spectrum sensing is a crucial task in a cognitive radio system. The transmissions of licensed users have to be really detected and spectrum sensing is thus the first step towards adaptive transmission in unused spectral bands without causing interference to primary users.

## 2. CENTRALIZED SPECTRUM SENSING

A cognitive radio is designed to be aware of and sensitive to the changes in its surrounding. The spectrum sensing function enables the cognitive radio to adapt to its environment by detecting spectrum holes. The most efficient way to detect spectrum holes is to detect the primary users that are receiving data within the communication range of user. In reality, however, it is difficult for a cognitive radio to have a direct measurement of a channel between a primary receiver and a transmitter. Thus, the most recent work focuses on primary transmitter detection based on local observations of users. The cognitive radio should distinguish between used and unused spectrum bands. Thus, the cognitive radio should have capability to determine if a signal from primary transmitter is locally present in a certain spectrum. Transmitter detection approach is based on the detection of the weak signal from a primary transmitter through the local observations of users. Basic hypothesis model for transmitter detection can be defined as follows [25]:

$$H_0 : x[n] = w[n]; \quad \text{signal is absent} \quad (1)$$

$$H_1 : x[n] = s[n] + w[n]; \quad \text{signal is present} \quad (2)$$

Where $n = 0, 1, 2, …, N\text{-}1$, $N$ is the sample index and $s[n]$ is the primary signal that is required to detect. The delay has not been taken into account. The null hypothesis $H_0$ states that no licensed user is present in the observed spectrum band. The alternative hypothesis $H_1$ indicates that some primary user signal exists. The centralized integrated spectrum sensing approach was considered which included the energy detection method for unknown narrow band signal and wavelet detection for wideband signal.

### 2.1. Energy Detection

If the primary user signal is a priori unknown to the secondary receiver, the optimal detector is an energy detector, also known as a radiometer [4]. Energy detection is the classical method for detecting unknown signals. It measures the energy of the received waveform over an observation time window [6]. First, the input signal is filtered with a bandpass filter (BPF) to select the bandwidth of interest. The filtered signal is squared and integrated over the observation interval. Finally, the output of the integrator is compared to a threshold to decide whether the primary user is present or not. When the spectral environment is analyzed in the digital domain, fast Fourier transform (FFT) based methods are usually used in order to obtain frequency response. FFT also generates the resolution in frequency domain. The periodogram method is a method to estimate power spectrum and it is based on the FFT [5]. The periodogram is a poor spectrum sensing method because of the large variance and bias of the estimate. Modified periodogram techniques have been developed to cope with such drawbacks [5], which is depicted in Figure 1.

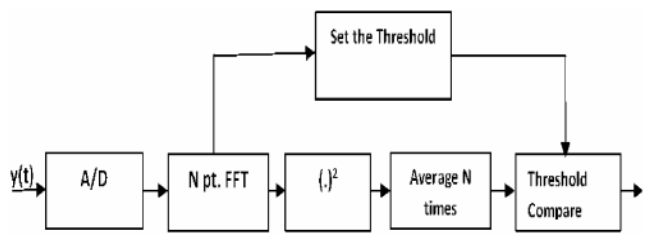

**Fig. 1: Digital Implementation of Wideband Energy Detector**

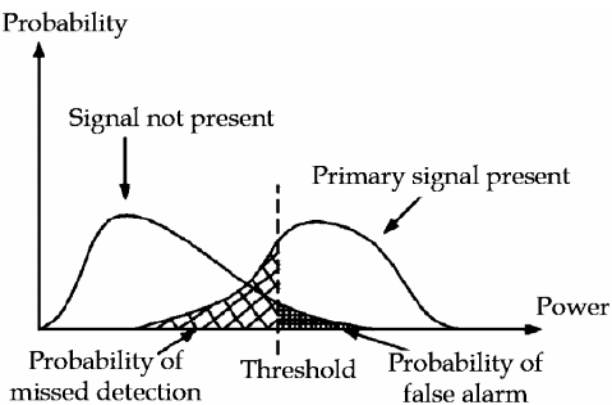

**Fig. 2: Trade-off between Missed Detections and False Alarms**

The performance of an energy detector is clearly degraded in fading environments and secondary users may need to cooperate in order to detect the presence of primary users [7]. One challenge is setting the right threshold for detection. The problem is illustrated in Figure 2 where the probability density functions of the received signal with and without primary signal are shown. When the probability of missed detections is very low, the probability of false alarms increases, resulting in low spectrum utilization. On the other hand, a low probability of false alarms would result in high missed detection probability, which increases the probability of interfering with the primary users. This trade-off has to be carefully considered. The threshold level is raised and lowered during detection to maintain a constant probability of a false alarm. This approach is known as constant false alarm rate (CFAR) detection.

## 2.2. Wavelet Detection

For the detection of wideband signals, the wavelet approach offers advantages in terms of both implementation cost and flexibility, in adapting to the dynamic spectrum as opposed to the conventional use of multiple narrowband bandpass filters (BPF). In order to identify the locations of vacant frequency bands, the entire wideband is modelled as a train of consecutive frequency sub-bands where the power spectral characteristic is smooth within each sub-band but changes abruptly on the border of two neighbouring sub-bands. By employing a wavelet transform of the power spectral density (PSD) of the observed signal x[n], the singularities of the PSD S(f) can be located and thus the vacant frequency bands can be found. One critical challenge of implementing the wavelet approach in practice is the high sampling rates for characterizing the large bandwidth.

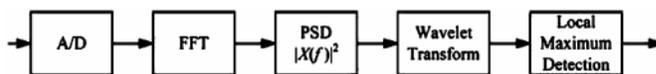

**Fig. 3: Digital Implementation of a Wavelet Detector**

The digital implementation of a wavelet detector for spectrum sensing is shown in Figure 3. When multiple systems co-exist in spectrum, it is difficult for the wavelet detector to identify the multiple systems because the wavelet detector is used to identify the locations of the vacant frequency bands and it is hard to cope with the inter-system interference environment.

## 3. RESULTS DISCUSSION

In the centralized spectrum sensing approach the signal were observed and the type of signal was identified using Welch's spectral estimation method and based on the this if it was narrow band then the energy detection technique was selected for the spectrum sensing for those specific signals otherwise the Wavelet detection techniques was selected. The integrated centralized approach has improved the spectral efficiency over the wide range of frequency. In the energy detector, the optimal threshold can be computed when the $P_{FA}$ is equal to the $P_{MD}$. Figure 4 shows that the probability of detection increase when $P_{FA}$ also in-crease. Moreover, the performance of energy detector is excellent when the SNR is greater than -5 dB. The result is shown in Figure 5. Again, if the probability of false alarm and probability of detection increase simultaneously and if the SNR greater that-5dB, regardless of the probability of false alarm the energy detector is optimum.

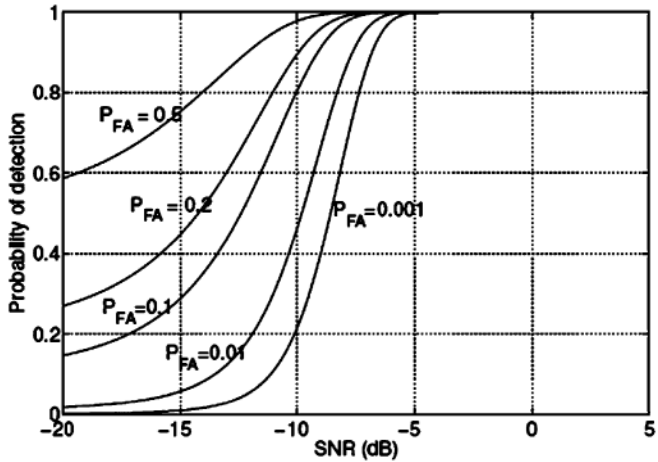

Fig. 4: Energy Detector: PD vs. SNR

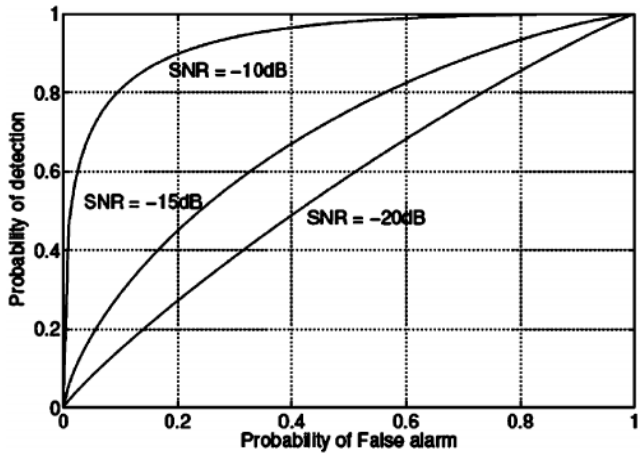

Fig. 5: Energy Detector: PD vs. PFA

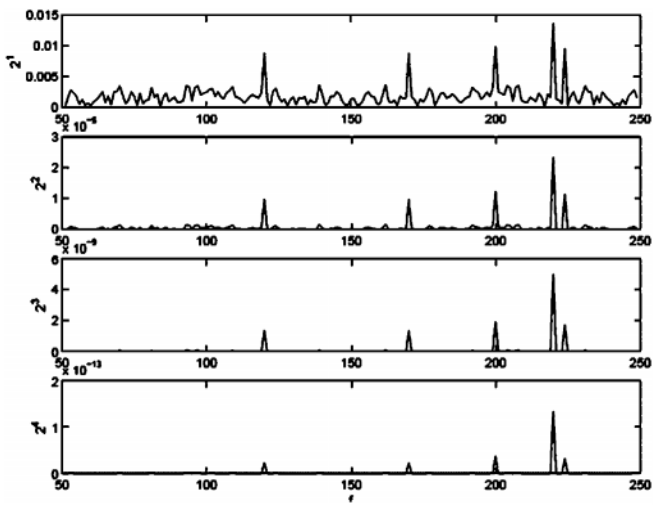

Fig. 6: Multiscale Wavelet Products of Signal

Figure 6 illustrates the multiscale products of wavelets expressed in [8]. Edges in the PSD Sr(f) are clearly captured by the wavelet transform in all curves. As the scale factor $s^j$ increases, the wavelet transform becomes smoother within each frequency band, retaining the lower-variation contour of the noisy PSD. In particular, the multiscale product method in Figure 6 is very effective in suppressing the spurious local extrema caused by noise, resulting in better detection and estimation performance.

## 4. CONCLUSION

In this paper the centralized integrated spectrum sensing approach investigated in the Multipath fading environment for unknown signals. The selection of the spectrum sensing technique was made on the basis of spectral width. The energy detection and wavelet detection techniques were employed and selection was made based on frequency band. The energy detection method performed better in narrow band frequency. The wavelet detection for wideband spectrum sensing required high sampling rates in order to characterize the entire wide bandwidth.